\documentclass[sn-mathphys-num]{sn-jnl}

\usepackage{graphicx}%
\usepackage{multirow}%
\usepackage{amsmath,amssymb,amsfonts}%
\usepackage{amsthm}%
\usepackage{mathrsfs}%
\usepackage[title]{appendix}%
\usepackage{xcolor}%
\usepackage{textcomp}%
\usepackage{manyfoot}%
\usepackage{booktabs}%
\usepackage{algorithm}%
\usepackage{algorithmicx}%
\usepackage{algpseudocode}%
\usepackage{listings}%
\usepackage{siunitx}%
\usepackage{makecell}
\usepackage{booktabs}  
\usepackage{float}     
\usepackage{arydshln}

\usepackage{braket}  
\usepackage[export]{adjustbox}

\usepackage{amsmath,amssymb}
\usepackage{graphicx}
\usepackage{physics}

\theoremstyle{thmstyleone}%
\theoremstyle{thmstyletwo}%

\theoremstyle{thmstylethree}%

\begin{document}
	
	\title[Long-distance DWDM quantum-network stabilization]{Exponential speedup of polarization stabilization for long distance DWDM quantum networks}


	\author*[1]{\fnm{Jinyi} \sur{Du}}\email{jinyidu@u.nus.edu}
	
	\author[1]{\fnm{En Teng} \sur{Lim}}
	
	\author[1,2]{\fnm{Xingjian} \sur{Zhang}}
	
	\author[3]{\fnm{Hongwei} \sur{Gao}}
	\author[3]{\fnm{George F.R.} \sur{Chen}}
	\author*[3]{\fnm{Dawn T. H.} \sur{Tan}}\email{dawn\_tan@sutd.edu.sg}
	\author*[1,4]{\fnm{Alexander} \sur{Ling}}\email{alexander.ling@nus.edu.sg}
	
	\affil[1]{\orgdiv{Centre for Quantum Technologies}, \orgname{National University of Singapore}, \orgaddress{\street{3 Science Drive 2}, \city{Singapore}, \postcode{117543}, \country{Singapore}}}
	
	\affil[2]{\orgdiv{Quantum Research Center}, \orgname{Technology Innovation Institute}, \orgaddress{\street{ Masdar City}, \city{Abu Dhabi}, \country{UAE}}}
	
	\affil[3]{\orgdiv{Photonics Devices and System Group}, \orgname{Singapore University of Technology and Design}, \orgaddress{\street{8 Somapah Rd}, \city{Singapore}, \postcode{487372}, \country{Singapore}}}

	\affil[4]{\orgdiv{Department of Physics}, \orgname{National University of Singapore}, \orgaddress{\street{2 Science Drive 3}, \city{Singapore}, \postcode{117551}, \country{Singapore}}}

	\abstract{
	Fibre-based quantum networks distributing polarization entanglement require a stable and uninterrupted transmission basis for reliable operation.
	Bright classical reference light enables rapid 	polarization feedback but can introduce noise into quantum channels. Entangled-photon-based feedback avoids this noise, but typically interrupts the target entanglement channel during calibration and becomes prohibitively slow
	over long distances due to the product loss of fibre links. Here we overcome both limitations by combining wavelength-bracketed probing with
	switch-enabled path decomposition. Spectrally adjacent entangled-photon sidebands track the polarization response of the central distribution channel	without interrupting its transmission, while optical switches and local	reference fibres independently determine the signal and idler network	transformations. Transferring the resulting compensation settings to the central channel eliminates calibration-induced downtime and changes the acquisition-time scaling from the product of the link losses to the sum of losses. We demonstrate the method on a 133~km fibre testbed and achieve continuous closed-loop stabilization for more than 24~hours without classical reference light. The decomposition of multi-link quantum feedback into single-link measurements provides a scalable stabilization strategy for wavelength-multiplexed quantum networks.
	}
	\keywords{DWDM quantum networks, entanglement distribution, polarization compensation}

\maketitle
\section{Introduction}\label{Introduction}

Distributing entanglement through optical fibre networks is a key enabling technology for quantum teleportation \cite{bennett1993teleporting,shen2023hertz,hu2023progress,liu2025chip,thomas2024quantum,takesue2015quantum}, entanglement-based quantum key distribution \cite{ekert1991quantum,bennett1992quantum,jiang2025entanglement,clark2025coexistence,neumann2022continuous,antesberger2024distribution,du2025entanglement,zhuang2025ultrabright}, and large-scale quantum information processing \cite{wehner2018quantum,kimble2008quantum,van2022entangling,knaut2024entanglement}.
Among the available photonic encodings, polarization is particularly
attractive because it can be readily manipulated and measured, and has
been widely deployed in both fibre
\cite{gul2026polarization,hubel2007high,wengerowsky2019entanglement}
and free-space links
\cite{ursin2007entanglement,yin2017satellite,ren2017ground,ma2012quantum,liu2021optical,amaral2025hybrid}.
However, single-mode fibres impose time-dependent polarization
transformations driven by temperature variations, bending, and mechanical
stress \cite{agrawal2010fiber}. For entangled photons distributed through
independent fibre links, these transformations cause the measurement
bases to drift
\cite{shi2021fibre,neumann2022continuous,zhou2025efficient}.
Polarization stabilization is therefore essential for
fibre-based entanglement distribution networks.

Polarization stabilization is commonly achieved using either classical
reference light or the distributed quantum photons themselves. Classical
reference schemes inject laser light into the fibre and obtain the channel
transformation from the measured reference polarization
\cite{heismann1994analysis,peranic2023study,martinelli2006polarization,thomas2024quantum,braband2025fast,gul2026polarization,amaral2025hybrid,craddock2024automated,chapman2024continuous,sena2025high,yin2025polarization,shi2026entanglement,zhang2025classical}.
Such schemes can provide rapid feedback, but introduce additional
coexistence constraints in dense wavelength-division-multiplexed (DWDM)
quantum networks. Photon leakage and Raman noise from the
reference laser can contaminate both the target and neighbouring quantum channels and saturate their single-photon detectors. Time multiplexing the classical reference and quantum signals
\cite{shi2026entanglement,gul2026polarization,gul2025noise,sena2025high,nardelli2026phase,zhang2025classical}
can suppress this contamination, but requires dedicated calibration
intervals during which quantum transmission is interrupted. For
dark-fibre DWDM networks intended to operate without auxiliary classical
pilot signals, it is therefore desirable to derive polarization feedback
directly from the distributed quantum photons.

Entangled-photon-based stabilization satisfies this requirement by using
the detected quantum correlations themselves as the feedback resource
\cite{dowling2023non,peranic2023study,shi2021fibre,shi2020stable,neumann2022continuous,jones2018tuning,du2025polarization,zhou2025efficient}.
QBER minimization
\cite{neumann2022continuous,jones2018tuning,shi2020stable},
hill-climbing optimization
\cite{dowling2023non,peranic2023study,du2025polarization},
gradient descent \cite{shi2021fibre}, and one-step state reconstruction
\cite{zhou2025efficient} have substantially reduced the number of
measurements required for polarization recovery. These developments
improve the algorithmic efficiency of the feedback procedure, but do not
change how the feedback photons are physically acquired.

Two practical limitations therefore still remain. First, polarization
recovery commonly requires dedicated measurements on the same wavelength
channels used for entanglement distribution. Calibration therefore
interrupts the actual entanglement transmission and
reduces the network duty cycle. Second, conventional two-link feedback
relies on coincidence events for which both photons have traversed their
respective lossy network fibres. The available feedback rate therefore
contains the product of the two link transmissions and decreases rapidly
as network loss increases. 

Here, we introduce a two-stage polarization stabilization method to address
both limitations through two successive forms of decoupling. Spectrally
adjacent entangled-photon sidebands decouple polarization feedback from the
central channel used for entanglement distribution, eliminating
calibration-induced transmission downtime. Local reference paths and optical
switching then decouple the two lossy network links within the feedback
measurement, replacing joint-link acquisition with independent single-link
measurements and thereby converting product-of-loss scaling into sum-of-loss
scaling. The resulting compensation settings are transferred to the central
entanglement channel. We demonstrate the complete method over fibre links of
87~km and 46~km, with respective losses of 18.5~dB and 10.5~dB. During more
than 24~hours of continuous closed-loop operation, the system maintains an
average central channel entanglement fidelity of 95.1\%. Together, these two
forms of decoupling provide a scalable route towards continuous, non-invasive
stabilization of entanglement distribution in DWDM quantum networks.

\section{Results}

\subsection{Wavelength-bracketed and switch-enabled stabilization method}
\label{sec:overall_concept}
Polarization-entangled photon pairs are generated through spontaneous
four wave mixing (SFWM) in a silicon photonic chip embedded in a Sagnac
interferometer \cite{du2025entanglement}. The broadband nonlinear process produces multiple frequency-conjugate signal-idler channel pairs satisfying
$2\omega_p=\omega_s+\omega_i$, where $\omega_p$, $\omega_s$, and $\omega_i$
denote the pump, signal, and idler frequencies, respectively. Each selected
signal-idler pair is prepared in the polarization Bell state
\begin{equation}
	\ket{\Phi^+}
	=
	\frac{\ket{H_sH_i}+\ket{V_sV_i}}{\sqrt{2}},
	\label{eq:bell_state_results}
\end{equation}
where $H$ and $V$ denote horizontal and vertical polarization, respectively. The broadband
spectrum provides multiple polarization-entangled channel pairs that can be
assigned different roles in the stabilization method. 

\begin{figure}[H]
	\centering
	\includegraphics[width=1.0\textwidth]{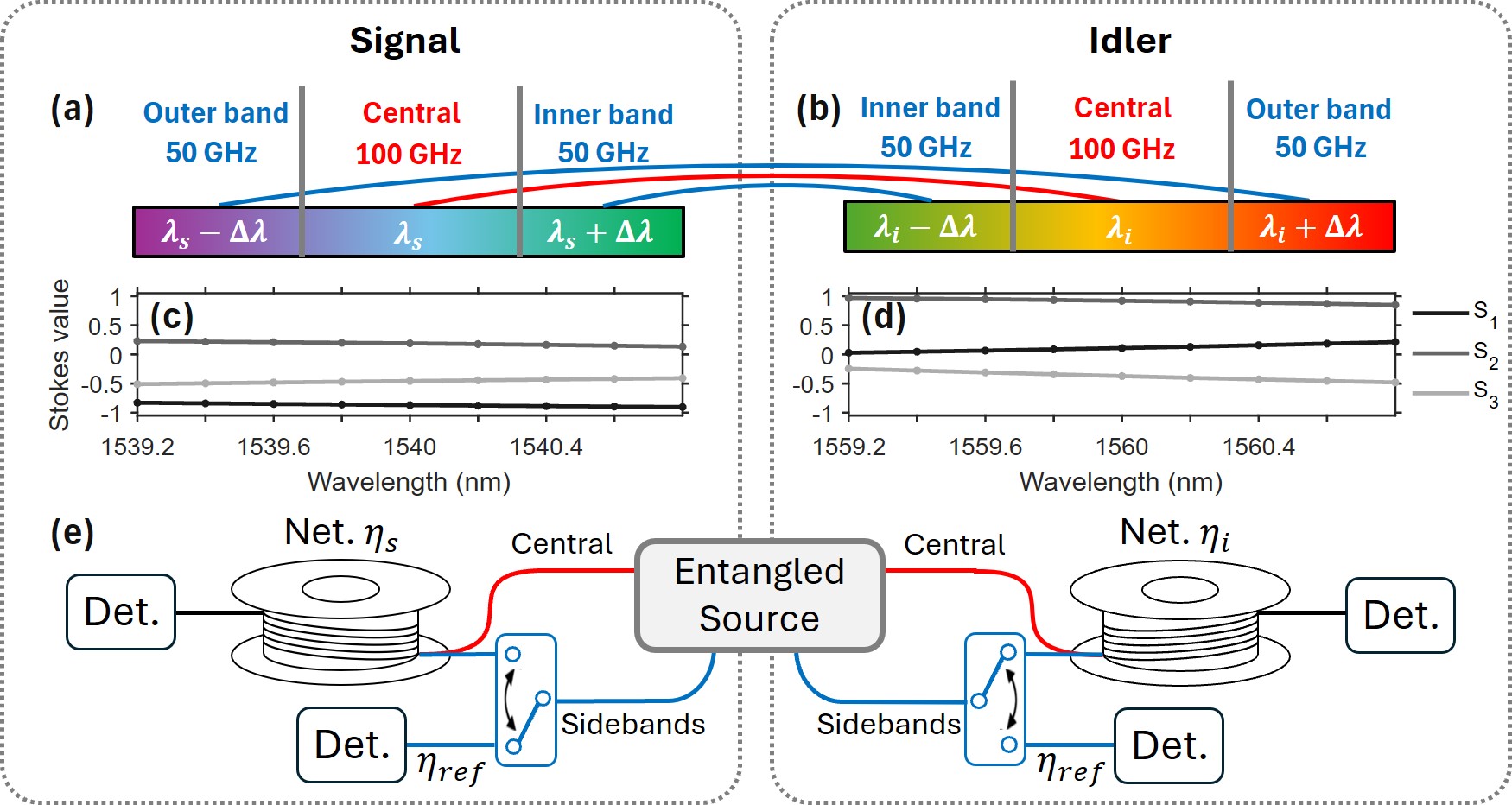}
	\caption{
		\textbf{Concept of wavelength-bracketed and switch-enabled polarization
			feedback.}
			The gray dashed enclosures indicate the signal (left) and idler (right) arms.
		\textbf{(a,b)} Spectral allocation of the signal and idler channels.
		The inner 50~GHz signal and idler channels form one entangled sideband pair,
		while the outer 50~GHz signal and idler channels form the other. The two
		sideband pairs bracket the central 100~GHz signal-idler pair at wavelength
		offsets $\pm\Delta\lambda$.
		\textbf{(c,d)} Smooth and continuous wavelength-dependent evolution of the
		Stokes vectors in the 87~km signal and 46~km idler network fibres.
		\textbf{(e)} Local-reference switching of the sidebands. The optical switches
		reconfigure only the paths of the sideband photons, while the central signal
		and idler channels remain continuously routed through their respective network (Net.)
		fibres. Detection (Det.) modules at the fibre outputs measure the
		entangled-photon correlations.
	}
	\label{fig:fig1}
\end{figure}

Figure~\ref{fig:fig1}~(a,b) illustrates the spectral allocation used for
wavelength-bracketed polarization feedback. The photons in outer signal and outer idler
channels are entangled, while the inner signal
and inner idler channels are also entangled. These two sideband pairs bracket a
central signal-idler pair, with wavelength offsets denoted by
$\Delta\lambda$ from the corresponding central signal and idler channels.

The polarization transformations introduced by the 87~km signal fibre and the 46~km idler fibre both vary continuously with wavelength, producing the smooth Stokes vector evolution
illustrated in Fig.~\ref{fig:fig1}~(c,d). Within a sufficiently narrow spectral
range, the polarization responses measured using the inner and outer sidebands
can therefore be averaged to estimate the response at the intermediate central
wavelengths. In the stabilization experiment, the sideband pairs are consequently
used as polarization probes, while the central pair is reserved for continuous
entanglement distribution. Polarization feedback can thus be acquired without
diverting the central channel for calibration.

However, wavelength-bracketed feedback remains subject to the loss penalty
associated with a two-link coincidence measurement. As shown in
Fig.~\ref{fig:fig1}~(e), when both photons of a sideband pair propagate through
their respective network fibres, a feedback coincidence is recorded only when
both photons survive the two lossy links. The coincidence rate therefore scales
as $\eta_s\eta_i$, where $\eta_s$ and $\eta_i$ denote the transmissions of the
signal and idler network fibres, respectively. The acquisition time for
conventional network-network ($\mathrm{NN}$) feedback consequently scales as
\begin{equation}
	T_{\mathrm{NN}}
	\propto
	\frac{1}{\eta_s\eta_i}.
	\label{eq:NN_acquisition_scaling}
\end{equation}

To decouple the two network-link losses within the feedback measurement, the
sidebands are switched between the lossy network fibres and short local
reference fibres. It is worth noting that only the sidebands are switched and the central signal and idler
photons remain routed through their respective network fibres throughout the
stabilization process. The short reference fibres provide stable, low-loss
polarization references that allow the polarization variations introduced by
the two network fibres to be measured independently.

In the first switched configuration, $\mathrm{R}_s\mathrm{N}_i$, the
polarization variation introduced by the idler network fibre is measured
against the signal reference fibre. Reversing the routing to
$\mathrm{N}_s\mathrm{R}_i$ similarly measures the polarization variation
introduced by the signal network fibre. This switching sequence provides
independent measurements of the polarization variations introduced by the two
network links.

Denoting the transmission of the local reference fibres by
$\eta_{\mathrm{ref}}$, the total acquisition time for the two switched
configurations scales as
\begin{equation}
	T_{\mathrm{switch}}
	\propto
	\frac{1}{\eta_{\mathrm{ref}}\eta_i}
	+
	\frac{1}{\eta_s\eta_{\mathrm{ref}}}
	\approx
	\frac{1}{\eta_i}
	+
	\frac{1}{\eta_s},
	\label{eq:switch_acquisition_scaling}
\end{equation}
where the propagation loss of the short local reference fibres is typically only 0.2~dB introduced mainly by the fibre connectors. This loss is negligible compared with the network fibre losses. The switching method therefore replaces the product-of-loss dependence of the joint network-network measurement with the sum-of-loss dependence of two single-link measurements.

\subsection{Wavelength-bracketed estimation of the central channel polarization}
\label{sec:wavelength-averaging}
\begin{figure}[H]
	\centering
	\includegraphics[width=1\textwidth]{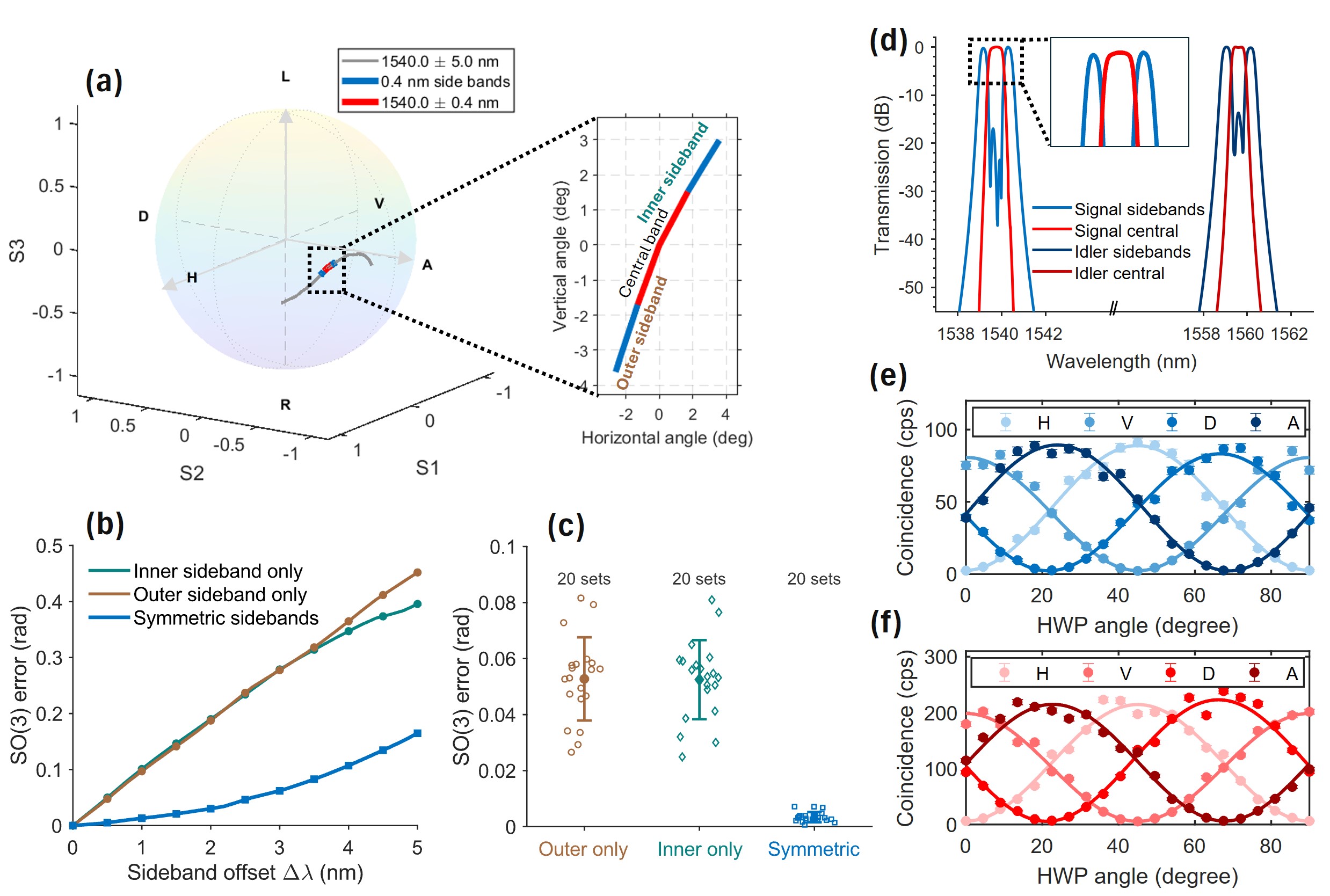}
	\caption{
		\textbf{Wavelength-bracketed estimation of the central-channel
			polarization response.}
		\textbf{(a)} Wavelength-dependent output polarization of the 87~km
		fibre spools, measured by scanning a fixed polarization laser over a
		$\pm5$~nm interval. The output states are represented on the Poincar\'e
		sphere, with the enlarged region showing their locally smooth evolution
		around the central wavelength.
		\textbf{(b)} $SO(3)$ angular error relative to the directly measured
		central channel rotation as a function of sideband offset. Symmetric
		bracketing produces a smaller error than either the inner or outer
		sideband alone.
		\textbf{(c)} $SO(3)$ angular errors obtained using the outer sideband,
		inner sideband, and symmetric sidebands across 20 independent experiments. Error bars represent
		one standard deviation.
		\textbf{(d)} Measured transmission spectra of filters used in the entangled-photon
		experiment. A pump at 1549.78~nm generates broadband signal-idler
		photon pairs. Each arm contains a 100~GHz central band, assigned to C47 (1539.77~nm)
		for the signal and C22 (1559.79~nm) for the idler, bracketed by two 50~GHz
		sidebands.
		\textbf{(e)} Two-fold coincidence fringes obtained from the symmetric
		sidebands.
		\textbf{(f)} Two-fold coincidence fringes measured in the central
		entanglement channel under the same fibre conditions as the sideband
		measurement.
	}
	\label{fig:wavelength-averaging}
\end{figure}

The wavelength-bracketing principle was first examined using a single input
polarization state transmitted through the 87~km fibre. Polarization-mode
dispersion (PMD) makes the fibre polarization transformation wavelength
dependent, causing the output state of polarization to evolve across the
spectrum. A tunable laser with a
fixed input polarization was scanned over a $\pm5$~nm interval around the
central wavelength $\lambda_s$~=~1540~nm, and the output Stokes vector was measured at
each wavelength. The resulting trajectory on the Poincar\'e sphere is shown in
Fig.~\ref{fig:wavelength-averaging}~(a). Although PMD produces substantial polarization evolution across the full $\pm5$~nm scan range, the enlarged region shows that its
wavelength dependence remains locally smooth over the narrower interval used
for sideband probing.

Under a first-order approximation, the Stokes vector of the central channel can therefore be estimated from the spherical midpoint of the two probe states at
$\lambda_s-\delta\lambda$ and $\lambda_s+\delta\lambda$. The symmetric bracketing cancels the first-order wavelength dependence, suppressing the estimation error compared with using a single neighboring spectral channel.

The suppression of the estimation error is quantified in
Fig.~\ref{fig:wavelength-averaging}~(b) as a function of the wavelength offset
between the sidebands and the central channel. The error is defined as the
three-dimensional rotation angle between the estimated and directly measured
central-channel polarization transformations in $SO(3)$. For either one-sided
probe, this angular error increases with the spectral distance from the central
wavelength. By contrast, averaging the symmetrically placed sidebands
consistently yields a smaller error across the measured range.
As shown in Fig.~\ref{fig:wavelength-averaging}~(c), the symmetric-sideband
estimate yields a mean $SO(3)$ angular error of $0.003\pm0.002$~rad across the
20 experiments, where the uncertainty denotes one standard
deviation. 

Having established the wavelength-bracketing accuracy for the fibre
polarization transformation, the method was next extended to the broadband
polarization-entangled photon source. The spectral filtering configuration
used for this measurement is shown in
Fig.~\ref{fig:wavelength-averaging}~(d). A pump laser at 1549.78~nm generated
broadband frequency-conjugate signal-idler photon pairs through the SFWM process. The broadband
output was separated into signal and idler arms, each containing a 100~GHz
central band bracketed by two neighbouring 50~GHz sidebands. The central
signal and idler bands were assigned to DWDM channels C47 (1539.77~nm) and C22 (1559.79~nm), respectively. The inner pair corresponds to the sidebands closer to the pump, whereas the outer pair corresponds to those farther from the pump.

Each filtered pair of signal and idler channels carries
polarization entanglement. The single photon polarization analysis was extended to the
two-photon state by measuring polarization correlations in the complementary horizontal (H), vertical (V), diagonal (D), and anti-diagonal (A) bases. For each spectral channel pair, the analyzer at signal arm was fixed at $H$, $V$, $D$, and $A$, while the other analyzer
was scanned to obtain the four corresponding two fold coincidence curves. This
measurement tested whether the sidebands reproduced the central-channel
polarization response across both measurement bases. Figure~\ref{fig:wavelength-averaging}~(e)
shows the two-fold coincidence measurement obtained from the symmetric
sideband channels with a fidelity of 97.3\%$\pm$0.3\%, while Fig.~\ref{fig:wavelength-averaging}~(f) shows the corresponding measurement of the central entanglement channel with a fidelity of 97.2\%$\pm$0.2\% under the same fibre condition. The two measurements exhibit closely matching
polarization correlations.
Thus, the compensation condition inferred from the symmetrically
bracketed sidebands provides an accurate correction to the central channel.

\begin{figure}[H]
	\centering
	\includegraphics[width=\textwidth]{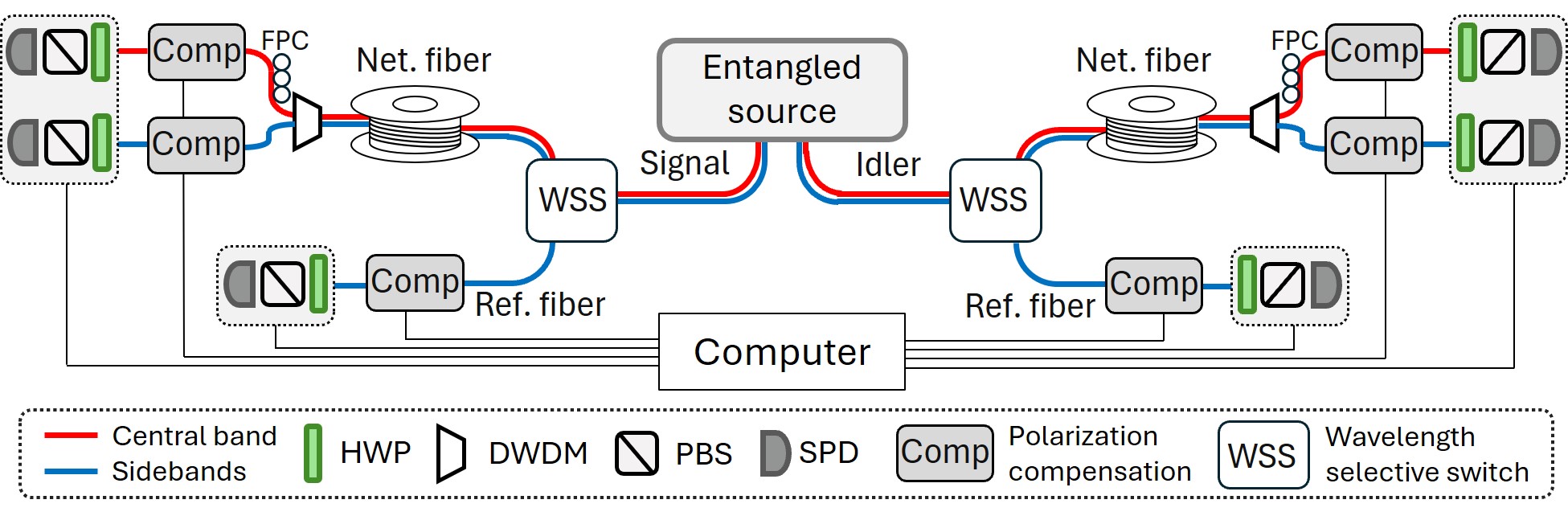}
	\caption{
		\textbf{Experimental setup for wavelength-bracketed and switch-enabled
			polarization stabilization.}
		Wavelength-selective switches route the sidebands between the network (N) and
		local reference (R) fibres, while the central signal and idler channels remain
		in the network paths. The network fibre in the signal arm is 87~km and 46~km in the idler arm. Polarization compensation and
		projection modules perform feedback optimization and entanglement measurements. WSS: wavelength selective switch; Comp.:	polarization compensation module; PBS: polarizing beam splitter; SPD: single photon detector; FPC: fibre polarization controller.
	}
	\label{fig:fullsetup}
\end{figure}

\subsection{Switch-enabled calibration and transfer of QHQ settings}
\label{sec:switch-protocol}
The detailed experimental implementation is shown in Fig.~\ref{fig:fullsetup}. A
wavelength-selective switch and a short local reference fibre are introduced
in each arm. The switches reroute only the sidebands between the network and
reference fibres, while the central signal and idler channels remain in their
respective network paths. At the end of the network fibre, the central and sideband
channels are spectrally separated and directed to independent polarization
compensation and projection modules. The compensation modules and polarization
analysers are computer controlled, enabling automated optimization from the
measured two-photon correlations.

\begin{figure}[H]
	\centering
	\includegraphics[width=\textwidth]{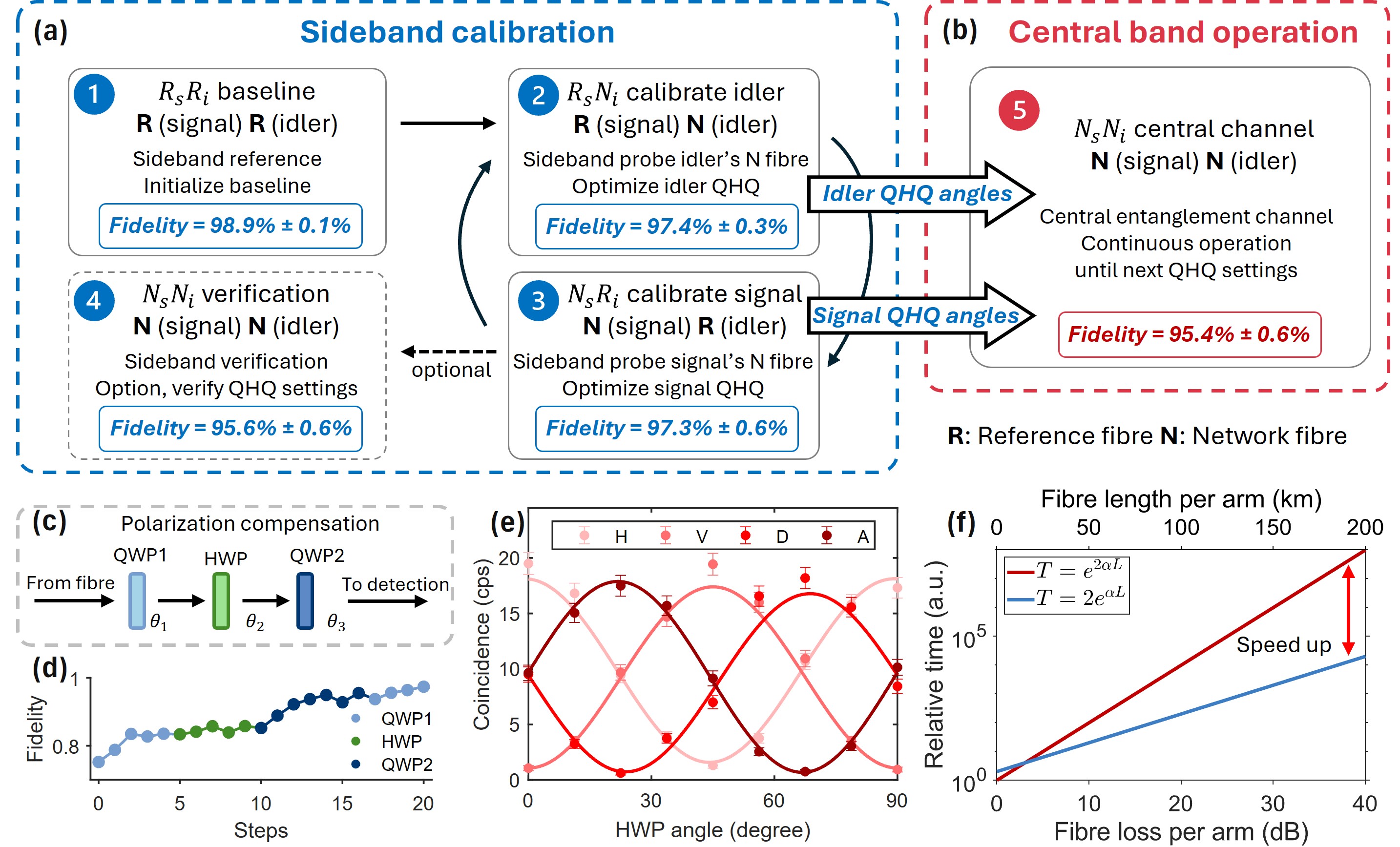}
	\caption{
		\textbf{Switch-enabled sideband calibration and transfer of QHQ settings
			to the central entanglement channel.}
		\textbf{(a)} Sideband calibration sequence. In Step~1, the
		$\mathrm{R}_s\mathrm{R}_i$ configuration establishes the local-reference
		baseline. In Step~2, $\mathrm{R}_s\mathrm{N}_i$ calibrates the idler
		network arm while the signal arm remains in the reference configuration.
		In Step~3, $\mathrm{N}_s\mathrm{R}_i$ calibrates the signal network arm
		while the idler arm remains in the reference configuration. For
		experimental verification, Step~4 applies the two independently determined
		QHQ settings simultaneously and evaluates them using an optional sideband
		$\mathrm{N}_s\mathrm{N}_i$ measurement.
		\textbf{(b)} In Step~5, the QHQ angles obtained from the sideband
		$\mathrm{R}_s\mathrm{N}_i$ and $\mathrm{N}_s\mathrm{R}_i$
		calibrations are transferred to the corresponding arms of the central
		entanglement channel, which remains continuously routed through the two
		network fibres.
		\textbf{(c)} Polarization compensation principle and QHQ module. The
		QHQ sequence is adjusted to approximate the inverse of the
		unitary polarization transformation introduced by the fibre.
		\textbf{(d)} Sequential hill climbing trajectory recorded
		during the Step~2 idler link calibration. The three wave plates are
		optimized successively, increasing the measured fidelity to 97.4\%$\pm$0.3\%.
		\textbf{(e)} Two-fold coincidence fringes measured in the central
		entanglement channel using the QHQ settings transferred from the sideband
		calibrations.
		\textbf{(f)} Acquisition time scaling for direct
		$\mathrm{N}_s\mathrm{N}_i$ feedback and switch-enabled calibration, showing the exponential acquisition time advantage of the
		switch-enabled method with increasing fibre length.
	}
	\label{fig:switch-protocol}
\end{figure}

The polarization transformation introduced by an optical fibre can be represented as a rotation of the Stokes vector, described by a matrix \(\mathcal{M}_{\mathrm{F}}\in SO(3)\). Polarization compensation is implemented using a QWP-HWP-QWP (QHQ) module placed after the fibre, as shown in Fig.~\ref{fig:switch-protocol}(c). The three independently adjustable waveplate angles allow the QHQ module to realize an arbitrary polarization rotation, represented by \(\mathcal{M}_{\mathrm{QHQ}}\). The QHQ angles are optimized such that this rotation compensates the fibre-induced transformation, giving
\begin{equation}
	\mathcal{M}_{\mathrm{QHQ}}\mathcal{M}_{\mathrm{F}} = I.
\end{equation}

The switch-enabled calibration sequence is summarized in
Fig.~\ref{fig:switch-protocol}~(a). Here, $\mathrm{R}$ and $\mathrm{N}$ denote
the local reference and network fibres, respectively, with the signal-arm
configuration listed first. In Step~1, fixed-polarization light is sent through the
two local reference fibres in the $\mathrm{R}_s\mathrm{R}_i$ configuration,
and the signal- and idler arm QHQ modules are adjusted to compensate the
polarization transformations introduced by the reference paths. The resulting
settings establish the reference baseline and are subsequently verified using
the sideband entangled photons, which yielded a fidelity of 98.9\%$\pm$0.1\%. The
reference arm QHQ settings are then fixed during the subsequent
network fibre calibrations.

In Step~2, the idler network fibre is calibrated in the
$\mathrm{R}_s\mathrm{N}_i$ configuration. The signal sideband propagates
through its reference fibre, while the idler sideband propagates through its
network fibre, and only the idler arm QHQ module is optimized using entangled photons. The QHQ
configuration is shown in Fig.~\ref{fig:switch-protocol}~(c). The three waveplates are optimized sequentially using a hill climbing procedure. Each waveplate is rotated stepwise in the direction that increases the measured
polarization-correlation fidelity. Once the fidelity passes a local maximum
and begins to decrease, the angle giving the highest fidelity is retained and
the optimization proceeds to the next waveplate. The trajectory shown in
Fig.~\ref{fig:switch-protocol}~(d) was recorded during Step~2 and reached a
final fidelity of 97.4\%$\pm$0.3\%.

In Step~3, the routing is reversed to
$\mathrm{N}_s\mathrm{R}_i$. The signal sideband propagates through its network
fibre, while the idler sideband remains in its reference path, and the same
hill climbing procedure is applied only to the signal arm QHQ module. This independent calibration of the signal link reached a fidelity of 97.3\%$\pm$0.6\%.
As the polarization transformations of the two local reference fibres remained
stable throughout the experiment, subsequent calibration cycles required only
Steps~2 and 3.

For experimental verification of the method, Step~4 applies the QHQ settings
obtained from the $\mathrm{R}_s\mathrm{N}_i$ and
$\mathrm{N}_s\mathrm{R}_i$ configurations simultaneously and evaluates them
using an additional sideband $\mathrm{N}_s\mathrm{N}_i$ measurement. The
resulting fidelity of 95.6\%$\pm$0.6\% confirmed that the two independently determined
settings remain valid when both network fibres are present. 
In continuous operation, this verification step can be omitted, and the method proceeds
directly from Steps~2 and 3 to Step~5. As shown in
Fig.~\ref{fig:switch-protocol}~(b), the same QHQ angles are transferred to the
corresponding arms of the central entanglement channel. The central channel,
which remains continuously routed through both network fibres, reached a
fidelity of 95.4\%$\pm$0.6\% using the transferred QHQ settings. 
The corresponding two-photon polarization correlation fringes are shown in Fig.~\ref{fig:switch-protocol}~(e).

The residual fidelity reduction in Steps~4 and 5 can be attributed to stronger PMD accumulated over the longer network fibres, which limits the achievable fidelity, together with a temporal mismatch between the sequentially obtained signal and idler QHQ settings caused by fibre drift during the hill climbing procedure.

In the present experiment, the signal and idler network fibre losses were
18.5~dB and 10.5~dB, respectively. The
$\mathrm{R}_s\mathrm{N}_i$ and $\mathrm{N}_s\mathrm{R}_i$ measurements each
required an integration time of 1~s, giving a total of 2~s to calibrate both
network links. By comparison, direct $\mathrm{N}_s\mathrm{N}_i$ feedback
required an integration time of 20~s to achieve the same photon count
requirement. The switch-enabled method therefore provided a tenfold reduction
in feedback acquisition time under the experimental link losses. 
More generally, as shown in Fig.~\ref{fig:switch-protocol}~(f), direct
$\mathrm{N}_s\mathrm{N}_i$ feedback follows a product-of-loss scaling, because
both photons must traverse their respective network links to contribute to the
feedback signal. By calibrating the two links independently, the switch-enabled
method converts this product-of-loss scaling into a sum-of-loss scaling. For
fibre transmission that decreases exponentially with distance, this difference
corresponds to an exponential reduction in the acquisition time, with the
advantage becoming increasingly pronounced at longer distances.

\subsection{Stable entanglement distribution over 24 hours}\label{sec:integrated-operation}
The wavelength-bracketed and switch-enabled methods were combined to stabilize
entanglement distribution over 87~km and 46~km signal and idler network fibres.
Throughout the experiment, the central signal and idler channels remained in
the $\mathrm{N}_s\mathrm{N}_i$ configuration, while the wavelength-selective
switches routed only the sideband probes between the network and reference
fibres. 

\begin{figure}[H]
	\centering
	\includegraphics[width=\textwidth]{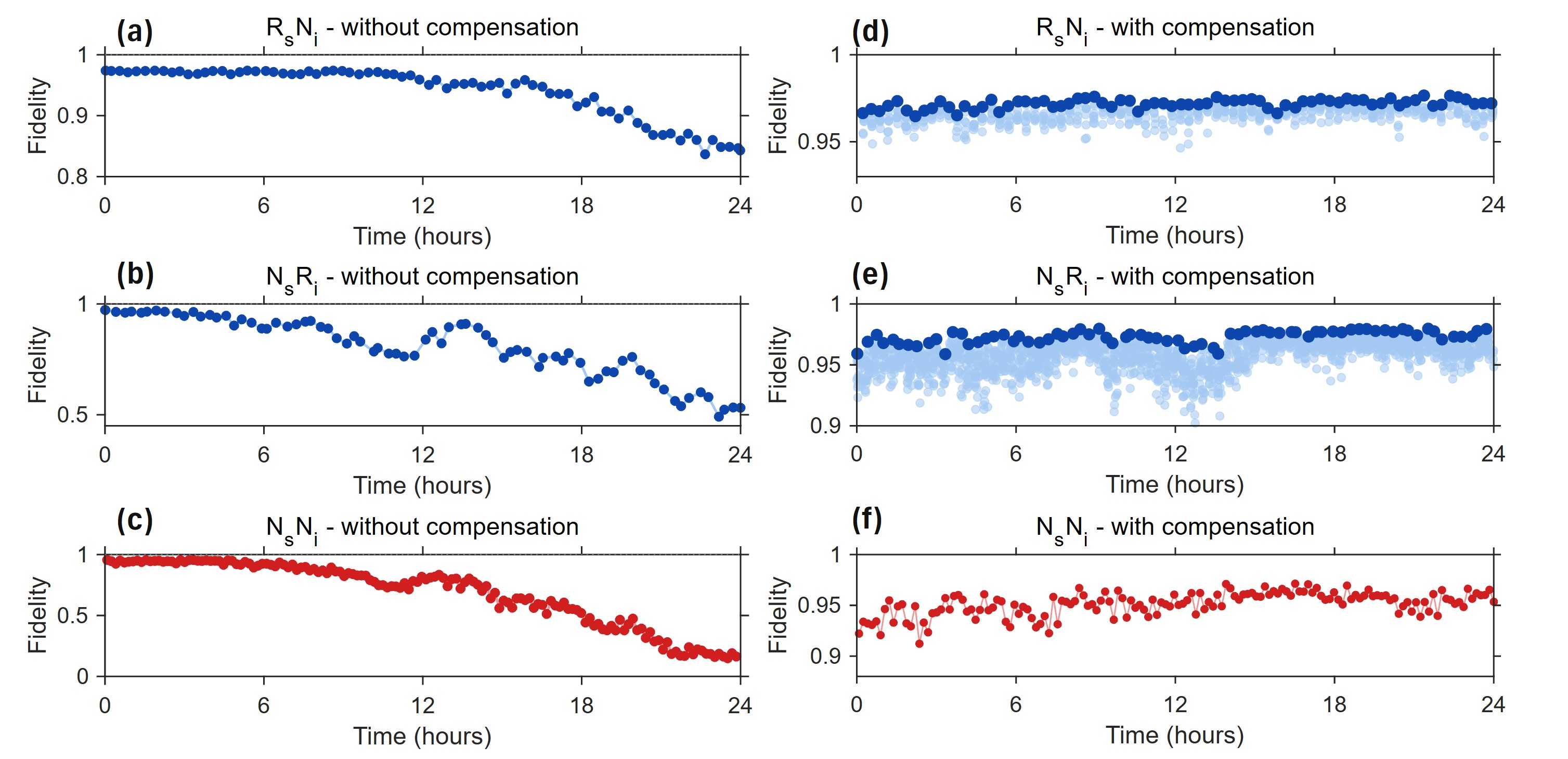}
	\caption{
		\textbf{Stable entanglement distribution over 24 hours.}
		\textbf{(a-c)} Polarization-correlation fidelities measured without
		active compensation in the $\mathrm{R}_s\mathrm{N}_i$,
		$\mathrm{N}_s\mathrm{R}_i$, and $\mathrm{N}_s\mathrm{N}_i$
		configurations, respectively. The fidelity decreases as the
		polarization transformations of the network fibres drift with time.
		\textbf{(d-f)} Corresponding fidelity records with active
		compensation. The $\mathrm{R}_s\mathrm{N}_i$ and
		$\mathrm{N}_s\mathrm{R}_i$ measurements independently track the idler
		and signal network links, respectively, while the
		$\mathrm{N}_s\mathrm{N}_i$ measurements record the stabilized central
		entanglement channel. In \textbf{(d,e)}, the lighter points show
		intermediate measurements acquired during QHQ optimization, and the
		darker points show the accepted fidelities.
	}
	\label{fig:integrated-operation}
\end{figure}

The long-term performance of the integrated stabilization method is shown in
Fig.~\ref{fig:integrated-operation}. Without active compensation, the
entanglement fidelities of all three fibre configurations drifted
substantially over 24~h as shown in Fig.~\ref{fig:integrated-operation}~(a-c). The
fidelity decreased in both the $\mathrm{R}_s\mathrm{N}_i$ and
$\mathrm{N}_s\mathrm{R}_i$ configurations as the polarization transformations
of the individual network fibres drifted with time. The largest degradation occurred in the
$\mathrm{N}_s\mathrm{N}_i$ configuration, where the time dependent
transformations of both network fibres contributed to the distributed
two photon state.

As shown in Fig.~\ref{fig:integrated-operation}~(d-f), active stabilization strongly suppressed these variations. The compensated
$\mathrm{R}_s\mathrm{N}_i$ and $\mathrm{N}_s\mathrm{R}_i$ measurements show
that the idler and signal network links were independently tracked throughout
the experiment. The corresponding QHQ settings stabilized the central
$\mathrm{N}_s\mathrm{N}_i$ channel, maintaining entanglement fidelity of 95.4\%$\pm$0.8\% over continuous 24~h operation, where the uncertainty denotes the temporal standard deviation. The lighter points in
Fig.~\ref{fig:integrated-operation}~(d,e) represent the intermediate
measurements acquired during QHQ optimization, whereas the darker points
indicate the accepted settings used for subsequent central channel
compensation.

The contrast between the uncompensated and compensated measurements
demonstrates that the wavelength-bracketed and switch-enabled feedback tracks the independent
polarization evolution of both network links and suppresses their combined
effect on the central entanglement channel. Stable entanglement distribution
is therefore maintained over the 87~km and 46~km network fibres without using
the central photons for polarization feedback.

\section{Discussion and conclusion}
This work demonstrates a polarization stabilization method for entanglement
distribution in DWDM fibre networks that does not require a copropagating
classical pilot during operation. The method combines wavelength-bracketed
estimation with switch-enabled link-wise calibration. The
wavelength-bracketed method uses two neighbouring sideband channels to
estimate the polarization correction required by the central entanglement
channel, allowing the central photons to remain continuously routed through
the network fibres and dedicated to entanglement distribution. 

The switch-enabled method independently determines the idler  and signal network
corrections in the $\mathrm{R}_s\mathrm{N}_i$ and
$\mathrm{N}_s\mathrm{R}_i$ configurations, respectively. For a fixed
coincidence count requirement, it replaces direct two-link feedback, whose
acquisition time scales as $1/(\eta_s\eta_i)$, with two single-link
measurements scaling as $1/\eta_s+1/\eta_i$. Under realistic link
losses, this reduced the feedback acquisition time with a tenfold improvement. The two methods enable polarization feedback without interrupting the central entanglement channel
while substantially reducing the required integration time. Experimentally,
the combined method maintained high fidelity entanglement distribution over
87~km and 46~km fibre links for more than 24~h.

The accuracy of wavelength-bracketed estimation depends on the PMD of the fibres. Symmetric bracketing
suppresses the leading wavelength dependent mismatch around the central
channel, whereas higher-order polarization mode dispersion produces a residual
error that increases with sideband separation. The usable probe spacing
therefore depends on the PMD of the deployed fibre and
must balance transfer accuracy against the photon flux available in the
sideband channels. For fibres with stronger higher order PMD,
the probe sidebands must be placed closer to the central channel.

Several aspects of the present implementation could be further improved.
First, the sequential hill climbing algorithm determines the QHQ settings by
rotating the three wave plates successively and evaluating the fidelity after
each adjustment. Although straightforward, this procedure requires many
intermediate measurements before reaching the optimum. More
measurement efficient reconstruction or matrix inversion methods could
determine the required compensation from fewer polarization projections,
thereby reducing the number of optimization steps
\cite{zhou2025efficient}. 

Second, the actuation speed is limited by the
mechanically rotated QWP-HWP-QWP modules. Although these modules provide the
degrees of freedom required to implement arbitrary polarization rotations,
each change in the compensation setting requires physical rotation of the wave
plates. Electro optic, liquid crystal or hybrid polarization controllers could
apply the calculated settings more rapidly and thereby reduce the actuation
time \cite{gul2026polarization,shi2021fibre,chin2025highly}. Improved
optimization algorithms would therefore reduce the number of settings that
must be measured, whereas faster polarization controllers would reduce the
time required to implement each setting.

A third opportunity lies in parallelizing the two network-fibre measurements. 
An additional pair of probe sidebands could be placed outside
the existing sideband pair. The
existing and additional sideband pairs could remain in the
$\mathrm{R}_s\mathrm{N}_i$ and $\mathrm{N}_s\mathrm{R}_i$
configurations, respectively. This would allow the idler  and signal network
transformations to be measured simultaneously. This parallel implementation
would remove the need to alternate between the two configurations and could
reduce the total acquisition time by up to a factor of two. The additional
sidebands would need to remain within the spectral range over which
the fibre polarization transformation can be transferred accurately to the
central channel.

The wavelength-bracketed and switch-enabled methods are specifically designed
for DWDM quantum networks. Broadband entangled photon sources naturally
provide multiple wavelength channels, allowing selected central channel pairs
to distribute entanglement while neighbouring sidebands provide polarization
feedback
\cite{joshi2020trusted,fan2025quantum,shi2025integrated,liu202240}. The
stabilization method can be incorporated directly into the spectral
architecture of a DWDM quantum network without interrupting its active
entanglement channels. This
use of neighbouring quantum channels for link-wise feedback makes the combined
method particularly suitable for continuous entanglement distribution in
lossy and wavelength-multiplexed quantum networks.

\section{Methods}
\subsection{Entangled photon source and single photon detection}

Polarization-entangled photon pairs were
generated by SFWM in a silicon photonic chip
embedded in a Sagnac interferometer. The source produced the Bell state
\begin{equation}
	\ket{\Phi^+}
	=
	\frac{\ket{H_sH_i}+\ket{V_sV_i}}{\sqrt{2}},
\end{equation}
where the subscripts $s$ and $i$ denote the signal and idler photons,
respectively. The broadband output was demultiplexed into frequency-conjugate
DWDM channel pairs.

After transmission through the selected fibre paths, the photons were directed
to polarization projection modules and detected using superconducting nanowire
single-photon detectors (SNSPDs) with system detection efficiencies of 55\%. The measured coincidence histogram peak had a full width at
half maximum (FWHM) of approximately 65~ps. A 50~ps coincidence window was used
throughout the experiment to suppress accidental coincidences and maintain a
high coincidence to accidental ratio (CAR)
\cite{du2024demonstration}.

Dispersion compensating fibre was included in each network link to limit
chromatic dispersion-induced broadening of the coincidence peak. The 87~km
signal network fibre comprised 79~km of SMF-28 fibre and 8~km of
dispersion compensating fibre, whereas the 46~km idler link comprised 42~km of
SMF-28 fibre and 4~km of dispersion compensating fibre. Without dispersion
compensation, the broadened coincidence peak would either reduce the fraction
of true coincidences captured by the 50~ps window or require a wider
coincidence window, which would degrade
the CAR and entanglement fidelity.

Long fibre transmission also introduces slow variations in the relative photon
arrival time, primarily through temperature induced changes in the optical
path lengths. Because a narrow coincidence window was used, the relative delay
was periodically recalibrated to prevent the coincidence peak from drifting
outside the selected window. For each recalibration, the coincidence histogram
was fitted to determine the peak position, and the electronic delay was updated
to realign the coincidence window with the fitted peak
\cite{du2025entanglement}.

\section*{Acknowledgements}
This project is supported by the National Research Foundation, Singapore through the National Quantum Office, hosted in A*STAR, under its Centre for Quantum Technologies Funding Initiative (S24Q2d0009).
D. T. H. Tan acknowledges funding from the National Research Foundation Investigatorship (NRF-NRFI08-2022-0003) and National Semiconductor Translation and Innovation Center (M24W1NS004 \& M24W1NS008).
The authors acknowledge insightful discussions from Dr. Hao Qin, Prof. Hoi-Kwong Lo, and Prof. Daniel Oi. ChatGPT was used for language polishing. All scientific content was reviewed and verified by the authors.

\section*{Author contributions}

J.D. conceived the wavelength-bracketed probing and switch-enabled path-decomposition method.
J.D. and X.Z. performed the simulations.
J.D. and E.T.L. performed the experiments.
H.G. and G.F.R.C. fabricated the photonic chips.
D.T.H.T. and A.L. supervised the project.
All authors discussed the results and contributed to the manuscript.

\section*{Competing interests}

The authors declare no competing interests.

\section*{Data availability}

The data supporting the findings of this study are available from the corresponding authors upon reasonable request.

\section*{Code availability}

The analysis code used in this study is available from the corresponding authors upon reasonable request.

\section{Supplementary Information}

\subsection{Stokes space formulation of wavelength-bracketed estimation}
\label{sec:wavelength-bracket-method}

For a narrow spectral channel, the polarization transformation of a
single mode fibre can be represented in Stokes space by a rotation
$\mathcal{M}(\lambda)\in\mathrm{SO}(3)$. The normalized input and output
Stokes vectors are related by

\begin{equation}
	\mathbf{s}_{\mathrm{out}}(\lambda)
	=
	\mathcal{M}(\lambda)\mathbf{s}_{\mathrm{in}},
	\qquad
	\mathcal{M}(\lambda)\in\mathrm{SO}(3),
	\label{eq:stokes_rotation_results}
\end{equation}

where the wavelength dependence of $\mathcal{M}(\lambda)$ arises from
polarization mode dispersion.

Let $\lambda_0$ denote the central channel wavelength. Two probe wavelengths
are placed symmetrically around it,

\begin{equation}
	\lambda_-=\lambda_0-\delta\lambda,
	\qquad
	\lambda_+=\lambda_0+\delta\lambda.
	\label{eq:bracket_wavelengths_results}
\end{equation}

When the two probe channels are detected jointly without spectral resolution,
their contributions to the measured Stokes vector are weighted by their
detected photon counts. Denoting these weights by $w_-$ and $w_+$, the
normalized bracketed output is

\begin{equation}
	\bar{\mathbf{s}}_{\mathrm{out,br}}
	=
	\frac{
		w_-\mathbf{s}_{\mathrm{out}}(\lambda_-)
		+
		w_+\mathbf{s}_{\mathrm{out}}(\lambda_+)
	}{
		w_-+w_+
	}.
	\label{eq:weighted_bracket_stokes}
\end{equation}

Using Eq.~\eqref{eq:stokes_rotation_results}, the bracketed response can be
written as

\begin{equation}
	\bar{\mathbf{s}}_{\mathrm{out,br}}
	=
	\bar{\mathcal{M}}_{\mathrm{br}}\mathbf{s}_{\mathrm{in}},
	\qquad
	\bar{\mathcal{M}}_{\mathrm{br}}
	=
	\frac{
		w_-\mathcal{M}(\lambda_-)
		+
		w_+\mathcal{M}(\lambda_+)
	}{
		w_-+w_+
	}.
	\label{eq:weighted_bracket_matrix}
\end{equation}

Expanding the two wavelength dependent rotations about $\lambda_0$ to first
order gives

\begin{equation}
	\mathcal{M}(\lambda_0\pm\delta\lambda)
	=
	\mathcal{M}(\lambda_0)
	\pm
	\delta\lambda\,\mathcal{M}'(\lambda_0)
	+
	O(\delta\lambda^2).
	\label{eq:taylor_stokes_rotation_results}
\end{equation}

Substitution into Eq.~\eqref{eq:weighted_bracket_matrix} yields

\begin{equation}
	\bar{\mathcal{M}}_{\mathrm{br}}
	=
	\mathcal{M}(\lambda_0)
	+
	\epsilon_w\delta\lambda\,
	\mathcal{M}'(\lambda_0)
	+
	O(\delta\lambda^2),
	\label{eq:weighted_bracket_expansion}
\end{equation}

where

\begin{equation}
	\epsilon_w
	=
	\frac{w_+-w_-}{w_++w_-}
	\label{eq:sideband_weight_imbalance}
\end{equation}

quantifies the imbalance between the detected contributions of the two probe
channels. Unequal weighting therefore leaves a first-order error proportional
to $\epsilon_w\delta\lambda$.

For balanced probe channels, $w_+=w_-$ and hence $\epsilon_w=0$. The
first order wavelength dependent terms cancel, giving

\begin{equation}
	\bar{\mathcal{M}}_{\mathrm{br}}
	=
	\frac{
		\mathcal{M}(\lambda_-)
		+
		\mathcal{M}(\lambda_+)
	}{2}
	=
	\mathcal{M}(\lambda_0)
	+
	O(\delta\lambda^2).
	\label{eq:balanced_bracket_average_results}
\end{equation}

Thus, symmetric and equally weighted wavelength bracketing eliminates the
first order wavelength dependent contribution, leaving a residual estimation
error that begins at second order in the probe separation.

\subsection{QHQ compensation and switch-enabled link decomposition}
\label{sec:stokes-switch-method}

The polarization state of a single photon can be represented by its normalized Stokes vector $\mathbf{s}$ through the Pauli expansion

\begin{equation}
	\rho
	=
	\frac{1}{2}
	\left(
	I+\mathbf{s}\cdot\boldsymbol{\sigma}
	\right),
	\label{eq:single_photon_stokes_density}
\end{equation}

where $\boldsymbol{\sigma}=(\sigma_x,\sigma_y,\sigma_z)$ denotes the Pauli matrices. A lossless polarization transformation corresponds to a rotation of the Stokes vector on the Poincar\'e sphere,

\begin{equation}
	\mathbf{s}_{\mathrm{out}}
	=
	\mathcal{M}\mathbf{s}_{\mathrm{in}},
	\qquad
	\mathcal{M}\in\mathrm{SO}(3).
	\label{eq:single_photon_stokes_rotation_switch}
\end{equation}

For a two-photon polarization state, the corresponding generalized Stokes parameters are

\begin{equation}
	S_{\mu\nu}
	=
	\mathrm{Tr}
	\left[
	\rho
	\left(
	\sigma_\mu\otimes\sigma_\nu
	\right)
	\right],
	\qquad
	\mu,\nu\in{0,x,y,z},
	\label{eq:generalized_stokes_parameters}
\end{equation}

where $\sigma_0=I$. The two-photon density matrix can therefore be reconstructed as

\begin{equation}
	\rho
	=
	\frac{1}{4}
	\sum_{\mu,\nu}
	S_{\mu\nu}
	\sigma_\mu\otimes\sigma_\nu.
	\label{eq:two_photon_stokes_density}
\end{equation}

For the input Bell state
$\ket{\Phi^+}=(\ket{HH}+\ket{VV})/\sqrt{2}$, the individual signal and idler Stokes vectors vanish, while the polarization correlations are described by the $3\times3$ correlation matrix

\begin{equation}
	T_0
	=
	\begin{pmatrix}
		1 & 0 & 0\\
		0 & -1 & 0\\
		0 & 0 & 1
	\end{pmatrix},
	\label{eq:bell_stokes_correlation}
\end{equation}

such that

\begin{equation}
	\rho_{\Phi^+}
	=
	\frac{1}{4}
	\left(
	I\otimes I
	+
	\sigma_x\otimes\sigma_x
	-
	\sigma_y\otimes\sigma_y
	+
	\sigma_z\otimes\sigma_z
	\right).
	\label{eq:bell_density_pauli}
\end{equation}

For each signal or idler arm, the polarization transformation introduced by the fibre is represented by a Stokes-space rotation
$\mathcal{M}_{a,\mathrm{F}}\in\mathrm{SO}(3)$, where
$a\in{s,i}$. Under independent polarization rotations in the two arms, the two-photon correlation matrix transforms as

\begin{equation}
	T_{\mathrm{F}}
	=
	\mathcal{M}_{s,\mathrm{F}}
	T_0
	\mathcal{M}_{i,\mathrm{F}}^{T}.
	\label{eq:two_photon_fibre_stokes}
\end{equation}

The QWP--HWP--QWP (QHQ) module implements a second Stokes-space rotation,

\begin{equation}
	\mathcal{M}_{a,\mathrm{C}}
	=
	\mathcal{M}_{\mathrm{QWP}}(\theta_{a,3})
	\mathcal{M}_{\mathrm{HWP}}(\theta_{a,2})
	\mathcal{M}_{\mathrm{QWP}}(\theta_{a,1}),
	\label{eq:QHQ_stokes_rotation}
\end{equation}

where $\theta_{a,1}$, $\theta_{a,2}$, and $\theta_{a,3}$ are the three waveplate angles. The QHQ module can realize an arbitrary polarization rotation and is adjusted to compensate the fibre transformation. Defining the total rotation after compensation as

\begin{equation}
	\mathcal{M}_{a,\mathrm{tot}}
	=
	\mathcal{M}_{a,\mathrm{C}}
	\mathcal{M}_{a,\mathrm{F}},
	\label{eq:combined_stokes_rotation}
\end{equation}

ideal compensation corresponds to

\begin{equation}
	\mathcal{M}_{a,\mathrm{tot}}
	=
	\mathcal{M}_{a,\mathrm{C}}
	\mathcal{M}_{a,\mathrm{F}}
	=
	I.
	\label{eq:single_arm_stokes_identity}
\end{equation}

After compensation in both arms, the Bell-state correlation matrix becomes

\begin{equation}
	T_{\mathrm{comp}}
	=
	\mathcal{M}_{s,\mathrm{tot}}
	T_0
	\mathcal{M}_{i,\mathrm{tot}}^T.
	\label{eq:general_compensated_stokes}
\end{equation}

When both arms satisfy Eq.~\eqref{eq:single_arm_stokes_identity},

\begin{equation}
	T_{\mathrm{comp}}
	=
	I T_0 I
	=
	T_0,
	\label{eq:compensated_bell_stokes}
\end{equation}

and the generalized Stokes parameters, and hence the two-photon density matrix, are restored to those of the input $\ket{\Phi^+}$ state.

To determine the network-fibre compensation settings independently, each arm contains a local reference path $\mathrm{R}$ and a network path $\mathrm{N}$. For arm $a$ and path $X\in{\mathrm{R},\mathrm{N}}$, the corresponding total rotation is defined as

\begin{equation}
	\mathcal{M}_{a,\mathrm{tot}}^X
	=
	\mathcal{M}_{a,\mathrm{C}}^X
	\mathcal{M}_{a,\mathrm{F}}^X.
	\label{eq:path_combined_rotation}
\end{equation}

For a general switched configuration $X_sY_i$, the measured polarization correlations are therefore described by

\begin{equation}
	T_{X_sY_i}
	=
	\mathcal{M}_{s,\mathrm{tot}}^X
	T_0
	\left(
	\mathcal{M}_{i,\mathrm{tot}}^Y
	\right)^T,
	\qquad
	X,Y\in{\mathrm{R},\mathrm{N}}.
	\label{eq:general_switched_stokes}
\end{equation}

The switch-enabled calibration follows the sequence

\begin{equation}
	\mathrm{R}_s\mathrm{R}_i
	\rightarrow
	\mathrm{R}_s\mathrm{N}_i
	\rightarrow
	\mathrm{N}_s\mathrm{R}_i
	\rightarrow
	\mathrm{N}_s\mathrm{N}_i.
	\label{eq:switch_sequence_stokes}
\end{equation}

In the first configuration, $\mathrm{R}_s\mathrm{R}_i$, the two reference paths are compensated such that

\begin{equation}
	\mathcal{M}_{s,\mathrm{tot}}^{\mathrm{R}}
	=
	I,
	\qquad
	\mathcal{M}_{i,\mathrm{tot}}^{\mathrm{R}}
	=
	I,
	\label{eq:RR_stokes_identity}
\end{equation}

and consequently

\begin{equation}
	T_{\mathrm{R}_s\mathrm{R}_i}
	=
	T_0.
	\label{eq:RR_stokes_result}
\end{equation}

In the $\mathrm{R}_s\mathrm{N}_i$ configuration, the signal photon remains in the compensated reference path, while only the idler network compensation is varied. Equation~\eqref{eq:general_switched_stokes} becomes

\begin{equation}
	T_{\mathrm{R}_s\mathrm{N}_i}
	=
	T_0
	\left(
	\mathcal{M}_{i,\mathrm{tot}}^{\mathrm{N}}
	\right)^T.
	\label{eq:RN_stokes}
\end{equation}

The idler QHQ setting is optimized until the Bell-state polarization correlations are restored,

\begin{equation}
	T_{\mathrm{R}_s\mathrm{N}_i}
	=
	T_0.
	\label{eq:RN_target_stokes}
\end{equation}

Because $T_0$ is invertible, Eqs.~\eqref{eq:RN_stokes} and \eqref{eq:RN_target_stokes} imply

\begin{equation}
	\mathcal{M}_{i,\mathrm{tot}}^{\mathrm{N}}
	=
	\mathcal{M}_{i,\mathrm{C}}^{\mathrm{N}}
	\mathcal{M}_{i,\mathrm{F}}^{\mathrm{N}}
	=
	I.
	\label{eq:idler_network_stokes_identity}
\end{equation}

Thus, the $\mathrm{R}_s\mathrm{N}_i$ measurement determines the idler network compensation independently of the signal network path.

Similarly, in the $\mathrm{N}_s\mathrm{R}_i$ configuration,

\begin{equation}
	T_{\mathrm{N}_s\mathrm{R}_i}
	=
	\mathcal{M}_{s,\mathrm{tot}}^{\mathrm{N}}
	T_0,
	\label{eq:NR_stokes}
\end{equation}

and restoration of the Bell-state correlations,

\begin{equation}
	T_{\mathrm{N}_s\mathrm{R}_i}
	=
	T_0,
\end{equation}

gives

\begin{equation}
	\mathcal{M}_{s,\mathrm{tot}}^{\mathrm{N}}
	=
	\mathcal{M}_{s,\mathrm{C}}^{\mathrm{N}}
	\mathcal{M}_{s,\mathrm{F}}^{\mathrm{N}}
	=
	I.
	\label{eq:signal_network_stokes_identity}
\end{equation}

After the signal and idler network compensations have been determined independently, the two settings are applied simultaneously in the $\mathrm{N}_s\mathrm{N}_i$ configuration. The resulting correlation matrix is

\begin{align}
	T_{\mathrm{N}_s\mathrm{N}_i}
	&=
	\mathcal{M}_{s,\mathrm{tot}}^{\mathrm{N}}
	T_0
	\left(
	\mathcal{M}_{i,\mathrm{tot}}^{\mathrm{N}}
	\right)^T
	\nonumber\\
	&=
	I T_0 I
	=
	T_0.
	\label{eq:NN_stokes_result}
\end{align}

The generalized Stokes parameters and the corresponding two-photon density matrix are therefore restored to those of the original Bell state. The $\mathrm{N}_s\mathrm{N}_i$ measurement is required only to verify the simultaneous operation of the two independently obtained compensation settings and is not required during their optimization. The switch-enabled method thus decomposes the original two-link stabilization problem into two independent single-link polarization-feedback problems.

\bibliography{sn-bibliography}

\end{document}